\documentstyle[amsfonts,twocolumn,prl,aps]{revtex}

\begin{document}

\twocolumn[
\hsize\textwidth\columnwidth\hsize\csname@twocolumnfalse\endcsname

\title{Laser generated wake fields as a new diagnostic tool for magnetized plasmas}
\author{Martin Servin and Gert Brodin}
\address{Department of Plasma Physics, Ume\aa\ University, S-901 87 Ume\aa , Sweden}
\maketitle

\begin{abstract}
In the presence of an external magnetic field, wake fields generated by a
short laser pulse can propagate out of the plasma, and thereby provide
information about the electron density profile. A method for reconstructing
the density profile from a measured wake field spectrum is proposed and a
numerical example is given. Finally, we compare our proposal with existing
plasma density diagnostic tecniques.
\end{abstract}

]
As is well-known, a short laser pulse propagating in an underdense
unmagnetized plasma can excite a wake field of plasma oscillations \cite
{Gorb}. This has interesting applications to plasma based particle
accelerators and is naturally of importance for the general understanding of
laser-plasma interactions. If the plasma is magnetized and the external
magnetic field non-parallel to the direction of propagation of the laser
pulse, the wake field becomes partially electromagnetic and thereby obtains
a nonzero group velocity \cite{Brodin}.

In the present paper we study wake field generation in an inhomogeneous
magnetized plasma. In particular we address the question to what extent the
wake field can propagate out of the plasma, and thereby provide information
about the plasma parameters. The most interesting case is that of a strongly
magnetized plasma (i.e. when the electron cyclotron frequency is larger than
the plasma frequency), for which almost all of the wake field energy -
except that generated in a narrow low density region - may propagate out of
the plasma. Since the wake field also in the magnetized case initially has
the frequency equal to the local plasma frequency, this provides a way of
extracting information about the electron density profile. It turns out that
even though wave overtaking -- for example when a higher frequency part of
the wake field passes a lower frequency part -- may occur, the density
profile can still be reconstructed by integrating the ray equations of
geometric optics backwards. This is more straightforward for a static
background density profile, and a numerical example is provided for this
case, where the predicted spectrum of the wake field corresponding to an
assumed density profile is shown, and a reconstructed profile is calculated.
Finally, the method is compared with two of the existing plasma density
diagnostic techniques \cite{Hartfuss}, namely interferometry and
reflectometry.

We consider a high frequency laser pulse with frequency $\omega _{H}$
propagating in a cold, weakly inhomogeneous magnetized plasma. We assume the
ordering $\omega _{H}\gg \omega _{p},\omega _{c}$, where $\omega _{p}$ and $%
\omega _{c}\equiv |q{\bf B}_{0}|/m$ are the plasma and electron cyclotron
frequency respectively, $q$ and $m$ are the electron charge and mass, and $%
{\bf B}_{0}=B_{0}\widehat{{\bf x}}$ is the (constant) external magnetic
field. \ We let the laser pulse propagate perpendicularly to the external
magnetic field. The ponderomotive force of the laser pulse will generate a
''low frequency'' wake field mode (which is the low frequency branch of the
extraordinary mode, or plasma oscillations modified by the magnetic field,
depending on the choice of terminology) during its path through the plasma.
The generation mechanism is most efficient if the pulse has a duration of
the order of the inverse plasma frequency or shorter. In principle, the
laser pulse will broaden due to ordinary dispersion, decrease its energy and
frequency due to the interaction with the wake field, etc. These and other
effects have been considered in homogeneous plasmas by for example Ref. \cite
{Brodin}. We will focus on the spectral properties of the wake field,
however, and for this purpose it turns out that we can forget about the
details of the laser pulse. Basically the effect of the laser field is to
provide a well localized ponderomotive source term in the governing
equations for the wake field, travelling with almost the speed of light in
vacuum.

The wake field quantities are denoted by index $L$. We introduce the
corresponding vector and scalar potentials ${\bf A}_{L}(z,t)$ and $\phi
_{L}(z,t)$, using Coulomb gauge, and the electron density is written $%
n=n_{0}(z,t)+n_{L}(z,t)$, where $n_{0}$ is the unperturbed density, assumed
to vary on longer space and time scales than the wake field. Furthermore,
the electron velocity ${\bf v}$ is divided into its high- and low frequency
part, and we denote the low frequency contributions perpendicular and
parallel to the direction of propagation with ${\bf v}_{L\perp }$ and{\bf \ }%
$v_{Lz}$ respectively. The ponderomotive force of the laser pulse induces
longitudinal wake field motion, which couple to motion in the $\widehat{{\bf %
y}}$ -direction through the Lorentz-force, but there is no wake field motion
in the direction of the external magnetic field, and accordingly we put $%
{\bf A}_{L}=A_{L}\widehat{{\bf y}}$ and ${\bf v}_{L\perp }=v_{L\perp }%
\widehat{{\bf y}}$. Linearizing in the low frequency variables, and
neglecting derivatives acting on $n_{0}$, we obtain the following set of
equations governing the wake field generation 
\begin{eqnarray}
-\mu _{0}qn_{0}v_{L\perp }+\left[ c^{-2}\partial _{t}^{2}-\partial _{z}^{2}%
\right] A_{L} &=&0  \label{lineq1} \\
c^{-2}\partial _{z}\partial _{t}\phi _{L}-\mu _{0}qn_{0}\left. v_{L}\right.
_{z} &=&0  \label{lineq2} \\
\partial _{t}v_{L\perp }-\omega _{c}v_{Lz}+\frac{q}{m}\partial _{t}A_{L} &=&0
\label{lineq3} \\
\partial _{t}v_{Lz}+\frac{q}{m}\partial _{z}\phi _{L}+\omega _{c}v_{L\perp }
&=&-\frac{q^{2}}{2m^{2}}\partial _{z}|{\bf A}_{H}|^{2}  \label{lineq4} \\
\partial _{t}n_{L}+n_{0}\partial _{z}v_{Lz} &=&0  \label{lineq5}
\end{eqnarray}
Next we write this on matrix form ${\Bbb A}{\bf u=b}$, where ${\bf u}=(\phi
_{L},n_{L},v_{Lz},v_{L\perp },A_{L})$, ${\Bbb A}$ is an operator matrix and $%
{\bf b}$ is the (ponderomotive) source vector. Since derivatives on $n_{0}$
are neglected, we can simplify the system (\ref{lineq1})-(\ref{lineq5}) by a
simple matrix inversion ${\bf u=}{\Bbb A}^{-1}{\bf b}$ treating the
operators in ${\Bbb A}$ like constant coefficients. We introduce the group
velocity of the laser pulse $v_{gH}\equiv \partial _{k}\omega $ and
eliminate differential operators in the denominators everywhere by
``multiplication''.  By using $\partial _{t}\approx -v_{gH}\partial
_{z}\approx -c\partial _{z}$ for derivatives acting on the {\em source vector%
}, and focusing on the first component of ${\bf u,}$ we finally obtain after
one space and one time integration 
\begin{equation}
\left[ \partial _{t}^{4}+(\omega _{h}^{2}+\omega _{p}^{2})\partial
_{t}^{2}-(\partial _{t}^{2}+\omega _{h}^{2})c^{2}\partial _{z}^{2}+\omega
_{p}^{4}\right] \phi _{L}=\frac{q\omega _{p}^{4}|{\bf A}_{H}^{2}|}{2m}
\label{waveeqs}
\end{equation}
where $\omega _{h}^{2}=\omega _{p}^{2}+\omega _{c}^{2}$ and we recognize the
operator acting on $\phi _{L}$ as the wave operator for the extraordinary
mode. Although derivatives on $n_{0}$ have been neglected in the derivation
of Eq. (\ref{waveeqs}), it should be emphasized that the equations for wake
field generation is more complicated here than for a homogeneous plasma.
This is because in the {\em inhomogeneous case}, the evolution of the wake
field is in general {\em not} quasi-static in a frame moving with the group
velocity of the laser pulse, and thus $\partial _{t}\approx -v_{gH}\partial
_{z}\approx -c\partial _{z}$ does not hold for derivatives acting on the
wake field.

It is useful to divide the study of the wake field properties into its
excitation and its propagation phase. The excitation of one additional
wavelength of the wake field takes place during a distance of the order of $%
2\pi c/\omega _{p}$, and -- as a basic assumption of ours -- the variations
of $n_{0}$ is negligible on this length scale. Thus as far as the excitation
process is concerned, the plasma can essentially be treated as homogeneous.
The solution for the wake field can thus be obtained from previous authors 
\cite{Brodin}. Changing to comoving coordinates $\xi =z-v_{gH}t$, $\tau =t$
the result for the wake field potential can be written 
\begin{equation}
\phi _{L}=\phi _{L0}\sin [k_{p}(\xi -\xi _{0})]  \label{initial}
\end{equation}
where $k_{p}$ is the wake field wavenumber $k_{p}=\omega _{p}/v_{gH}$, $\xi
_{0}$ is the (constant) position of the (short) laser pulse, and $\phi
_{L0}=(q\omega _{p}/mv_{g})\int_{-\infty }^{\infty }|{\bf A}_{H}|^{2}d\xi $.
The important result here, for our purposes, is the determination of the
initial value of the wake field wave number $k_{p}=\omega _{p}/v_{gH}$,
which corresponds to an initial frequency $\omega _{p}$ (in the laboratory
frame), that will vary with the position of generation.

Next we make the ansatz of geometric optics. Inspecting the wave operator in
Eq. (\ref{waveeqs}), we note that there is a resonance at $\omega
^{2}=\omega _{h}^{2}\equiv $\ $\omega _{c}^{2}+\omega _{p}^{2}$\ and
cut-offs at $\omega _{L}\equiv $ $\frac{1}{2}[-\omega _{c}+(\omega
_{c}^{2}+4\omega _{p}^{2})^{1/2}]$ and $\omega _{R}\equiv $ $\frac{1}{2}%
[\omega _{c}+(\omega _{c}^{2}+4\omega _{p}^{2})^{1/2}]$. The dispersion
relation has two positive roots. One branch is valid for $\omega >\omega
_{R} $ and one for $\omega _{L}<\omega <\omega _{h}$. As the wake field is
generated with the local plasma frequency $\omega _{p},$ the wake field must
belong to the latter branch. Therefore we write the dispersion relation from
now on as 
\begin{equation}
\omega =W(k,z,t)\equiv \left\{ \chi -\sqrt{\chi ^{2}-\omega
_{h}^{2}c^{2}k^{2}-\omega _{p}^{4}}\right\} ^{1/2}  \label{dr}
\end{equation}
where $\chi \equiv (\omega _{h}^{2}+\omega _{p}^{2}+c^{2}k^{2})/2$. Due to
the cut-off $\omega _{L}$ and the resonance $\omega _{h}$, only parts of the
wake field satisfying the inequality $\omega _{L}<\omega <\omega _{h}$ at
all points on its path through the plasma can leave the plasma and
contribute to the wake field spectrum. Within the geometric optics
approximation, the path of different parts of the wake field is determined
by the {\em ray equations} \cite{Whitham} 
\begin{equation}
\frac{dk}{dt}=-\partial _{z}W\text{ , \ \ \ \ }\frac{d\omega }{dt}=\partial
_{t}W\text{\ \ \ }  \label{ray}
\end{equation}
where $d/dt\equiv \partial _{t}+v_{g}\partial _{z}$. Specifically, for a
time-independent medium, the right hand side of the last equation is zero
and the wake field propagates, with the local wake field group velocity $%
v_{g}\equiv \partial _{k}\omega $ with unchanged frequency.

\label{reconstr}It is natural to first consider the density profile
reconstruction from a wake field in the case of a quasi-static plasma. The
proposed technique can be applied not only to truly static density profiles
but to the large class of phenomenons varying on a time scale significantly
larger than the time of propagation of the wake field through the plasma
slab. Then, sequential laser pulses gives a sequence of wake fields spectra
from which a ``movie'' of the density profile can be obtained.

We assume that the generated wake field spectrum is measured immediately
outside the plasma boundary. In consistence with the geometric optics
approximation we will treat the measured data as a weakly time dependent
spectrum with well defined sharp (quasi-monochromatic) peaks. The data is
not necessarily monochromatic at a given time, however. A part of the wake
field generated at an earlier time may overtake some part of the wake field
generated at a later time, leading to a spectrum with multiple sharp peaks.
Generally, we can express the data as a set of distinct frequencies measured
at different times, in which case a sharp curve can be recognized, see
Fig.1(a). Due to cut-offs and/or resonances, the curve may be discontinuous,
and the algorithm below only works until the first jump. 

Before presenting the reconstruction algorithm we need to introduce some
notations. As mentioned, the measurement results in a function of time, in
general multivalued, describing a path $\omega (t)$ as illustrated in
Fig.1(a). Let the path be discretized into consecutive points numbered $%
i=0,1,...,n$, $i=0$ being the first measured point. Let also the plasma,
having length $L$, be discretized into cells, numbered $j=1,2,...,N$ , of
length $\Delta z=L/N$ so that the plasma has the plasma frequency $\omega
_{p(j)}$ at $(j-1)\Delta z\leq z\leq j\Delta z$. In order to distinguish the
two discretizations we will use upper indices for quantities discretized
according to the spectrum-path discretization, e.g. $t^{(i)}$, and lower
indices for quantities discretized according to the plasma discretization,
e.g. $\omega _{p(j)}$. In particular we will let $z_{(j)}\equiv (j-1)\Delta z
$ whereas $z^{(i)}$ is the ``exact'' point in the plasma where the wake
field corresponding to the $i$:th point in the spectrum was generated.
Furthermore, it is convenient to introduce $\Delta z^{(i)}$, which is
defined in Fig.1(b).

The following algorithm can be used to retrace the measured frequency
spectrum and thereby reconstruct the electron density profile: We assume
that the points $0,1,2,...,i-1$ (grey dots in Fig.1(b)) have already been
retraced so that the plasma frequency of the cells $j+1,j+2,...,N$ (black)
is known and the plasma frequency of cell $j$ (grey) has yet to be
determined. Consider now the retracing of the next spectrum-point $i$,
corresponding to the frequency $\omega ^{(i)}$ detected at time $t^{(i)}$.
Given the plasma frequency of the already reconstructed cells, the path of
spectrum point $i$ is known up to cell $j$, where the plasma frequency and
thereby the group velocity is yet to be determined. Since spectrum point $i$
at least belongs in the neigborhood of cell $j,$ a first guess -- correct to
lowest order -- is that $\omega _{p(j)}=\omega ^{(i)}$. Making this guess,
the path of spectrum point $i$ can be reconstructed a bit further by
comparing the detection time, $t^{(i)}$, with the generation time of this
particular wake field, and using the ray equations (\ref{ray}). In
particular it can be decided whether this point belongs to cell $j$ or cell $%
j-1$. If it belongs to cell $\ j$, the algorithm assigns the calculated
position to point $i$ and proceeds to spectrum point $i+1$. If it belongs to 
$j-1$, the cell $j$ is filled with spectrum points and a definite plasma
frequency is given to this cell, before the algorithm continues in the same
manner with reconstructing cell $j-1$ and the path of spectrum point $i+1.$
For density variations on a faster time-scale than $c/L$, the above
algorithm does not work. Preliminary work suggests that it can be
generalized to a more rapidly varying regime, however, but it would require
multiple laser pulses propagating in the plasma slab simultaneously.

In order to demonstrate the method we numerically calculate a wake field
spectrum using the ray equations (\ref{ray}) starting from an assumed
density profile, see Fig.2(a). In the reconstruction algorithm this spectrum
is then treated as experimental data. In this case we consider a plasma
magnetized such that $\omega _{c}=1.1\times \omega _{p,\text{max}}$, where $%
\omega _{p,\text{max}}$ is the maximum value of the plasma frequency. For
simplicity we normalize such that $L=1$, $c=1$, and we let the laser pulse
enter the plasma at $t=0$ and exit at $t=1$. The spectrum is obtained by
treating the wake field as a collection of point particles -- jumping into
existence as the laser pulse propagates through the plasma -- each moving
with the local group velocity and the initial wavenumber and frequency
determined by Eq. (\ref{initial}). Since the density profile is known, in
contrast to the case of retracing the spectrum, it is straight forward to
propagate these ``particles'' by means of the ray equations (\ref{ray}).

Retracing the spectrum according to the algorithm presented above results in
a density profile that can be compared with the one we assumed, see Fig.
2(b). A small numerical error -- that can be removed with a finer
discretization -- can be seen. Note that the entire plasma profile cannot be
reconstructed. The left most points in Fig. 2(b) are missing. This is
because the wake field generated in this region of low density cannot
propagate through the plasma since there is a cut-off prohibiting this. The
information of this region is already missing in the wake field spectrum.

Existing techniques for unperturbing plasma density diagnostics include
reflectometry and interferometry methods as well as incoherent scattering 
\cite{Hartfuss}. It is most natural to compare our proposal with the two
former methods: Reflectometry essentially makes use of the reflection of
electromagnetic waves against a cut-off occurring at a point where the
frequency of the electromagnetic wave equals the local plasma frequency. The
plasma frequency as a function of position can be derived from
time-of-flight measurements for a suitable spectrum of electromagnetic
waves. Applying interferometry, a laser beam is split-up into two (or more)
beams. One of the beams passes through the plasma and experiences a phase
shift relative to the beam propagating in vacuum. This phase shift --
obtained by letting the beams interfere -- is related to the density {\em %
averaged along the path }through the plasma.

In many respects our new method compare favorably with existing techniques,
provided the basic criterion of a strong external magnetic field is
fulfilled. Some advantages are: 1){\em \ Coaxial profile information: }%
Interferometry can at best provide information about the density profile in
a direction {\em perpendicular} to the penetrating laser beams (if multiple
beams are used) whereas the proposed method reconstructs the density profile 
{\em along} the path of the laser pulse. 2){\em \ High spatial resolution: }%
The requirement that the length scales of inhomogeneities must be much
larger than the local plasma wave length $\lambda _{p}\equiv 2\pi
v_{g}/\omega _{p}$ sets the spatial resolution limit to roughly $%
10^{8}n^{-1/2}{\rm m}$ for a strongly magnetized plasma ($\omega _{c}\sim
\omega _{p})$, where $n$ is the electron number density in units {\rm m}$%
^{-3}$. 3){\em \ High temporal resolution:} The temporal resolution limit is
equal to the time that separates sequential laser pulses probing the
dynamics, and thus $t_{{\rm res}}\sim \Delta L/c,$ where $\Delta L$ is the \
spatial separation of the laser pulses. For use of the ''static'' density
profile reconstruction scheme $\Delta L$ must be comparable to the width of
the plasma column, implying that $t_{{\rm res}}$ is not longer than $10^{-8}%
{\rm s}$, for typical laboratory plasma dimensions. 4){\em \ Density minimum
access } In our method the density profile can be measured even in the
vicinity of a local density minimum, in contrast to reflectometric methods
for which only monotonous profiles can be fully reconstructed. This feature
together with points 2) and 3) makes it possible to study the density
evolution in detail of various dynamical processes, e.g. wave propagation
and instabilities, using our method.

Given a suitable plasma -- a regime close to thermonuclear fusion conditions
seems appropriate -- the equipment necessary to apply the proposed scheme is
firstly a laser that can produce powerful and short pulses. Commercial
Ti:sapphire lasers typically fulfill these requirements by far. Secondly,
the detector need to operate over a broad frequency band and produce
spectrums that are time-resolved on (roughly) the time-scale $L_{{\rm ih}}/c$%
, where $L_{{\rm ih}}$ is the density inhomogeneity scale length. Similar
detector requirements have been fulfilled in the applications of ultra-short
pulse reflectometry \cite{Cohen}, however, and thus we deduce that it is 
{\em not} necessary to develop new technology for our scheme to work.

There are a numbers of issues not addressed in our density reconstruction
scheme. Firstly, what happens if the external magnetic field is not
homogeneous? This is simple provided the magnetic field profile is know,
since our static reconstruction scheme is essentially unaffected by this
change, and we only need to include the corresponding variations in the
dispersion relation (\ref{dr}). Secondly, what happens if the profile to be
reconstructed has two- or three-dimensional spatial variations?
Unfortunately -- unless the extra variations occur on longer spatial scales
-- this complicates the application a lot, at least from a practical point
of view. Two parts of the wake field generated at different positions will
follow different orbits in the plasma and exit at different locations. By
using multiple lasers and detectors placed at suitable positions, it seems
possible that 3-D density profiles can be reconstructed in principle, but to
address this question in any detail is beyond the scope of this paper.
Finally we point out that the question of the usefulness of our scheme is
unlikely to be settled by purely theoretical arguments, and instead we want
to encourage experimental investigations of wake field generation in
magnetized plasmas. A first interesting step, before complete density
profile reconstructions are performed, would be to experimentally verify
that the maximum frequency of the wake field coincide with the maximum value
of the local plasma frequency.

\bigskip \newpage

\section{Figure captions}

Fig. 1 Illustration of the discretization scheme. (a) Spectrum path
discretization (b) Plasma slab discretization

Fig. 2 Example of a reconstruction. (a) Predicted wake field spectrum
normalized against the maximum frequency (b) . Normalized plasma profiles.
The solid line marks the assumed profile and the crosses marks the
reconstructed profile.


\begin{references}
\bibitem{Gorb}  T. Tajima and J. M. Dawson, Phys. Rev. Lett. A {\bf 43}, 276
(1979); L. M. Gorbunov and V. I. Kirsanov, Zh.Eksp. Theor. Fiz. {\bf 93},
509 (1987). [Sov.Phys. JETP {\bf 66}, 290 (1987)]; J. M. Dawson, Phys. Scr. 
{\bf T 52}, 7 (1994).

\bibitem{Brodin}  G. Brodin and Jonas Lundberg, Phys. Rev. E {\bf 57}, 7041
(1998).

\bibitem{Hartfuss}  H. J. Hartfuss, Plasma Phys. Control. Fusion {\bf 40}
A231-A250 (1998); I. H. Hutchinson, {\em Principles of plasma diagnostics}
(Cambridge University Press, Cambridge, England, 1987).

\bibitem{Cohen}  B. I. Cohen, E. B. Hooper, T. B. Kaiser, E. A. Williams and
C. W. Domier, Phys. Plasmas, {\bf 6, }1732 (1999).

\bibitem{Whitham}  G. B Whitham, {\em Linear and Nonlinear Waves }(John
Wiley \& Sons, New York, 1974)
\end{references}
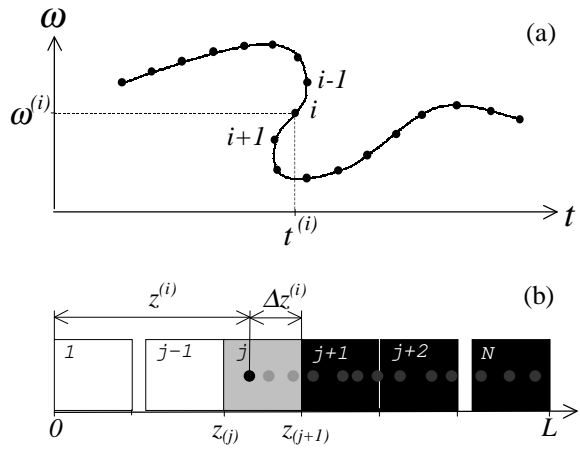
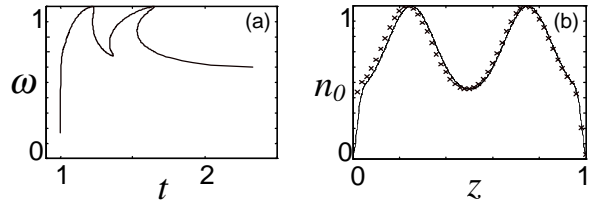
\end{document}